\documentstyle[12pt]{article}
\newcommand{\be}{\begin{equation}}
\newcommand{\ee}{\end{equation}}
\newcommand{\beqa}{\begin{equationarray}}
\newcommand{\eeqa}{\end{equationarray}}
\newcommand{\bibi}{\bibitem}

\begin{document}
\begin{flushright}
July 20, 1996\\
\end{flushright}
\begin{center}
\Large \bf A Decrumpling Model of the Universe  
\end{center}
\vspace{0.5 cm}  
\begin{center}
\Large {M. Khorrami}$^{1,2}$, 
   { R. Mansouri}$^{3,4,*}$, { M. Mohazzab}$^5$, 
\end{center} 
\vskip 0.5cm
{\it

\vskip -0.5cm
\noindent $^1$ Department of Physics, Tehran University,
             North-Kargar Ave. Tehran, Iran.\\ 

\vskip -0.5cm
\noindent $^2$ Institute for Advanced Studies in Basic Sciences,
             P.O.Box 159, Gava Zang, Zanjan 45195, Iran.\\ 

\vskip -0.5cm
\noindent $^3$ Department of Physics, Sharif University of Technology,
             P.O.Box 9161, Tehran 11365, Iran.\\ 

\vskip -0.5cm  
\noindent $^4$ Universitaet Potsdam, Mathematisches Institut, Kosmologie 
Gruppe, Potsdam, Germany.\\  

\vskip -0.5cm
\noindent $^5$ Physics Department, Brown University, Providence RI. 02912,
 USA.\\ 
\noindent $^*$ E-Mail: rmansouri@aip.de}

\vskip 0.5cm

\noindent \small {\bf Abstract}: Assuming a cellular structure for the 
space-time, we propose a model in which the expansion of the universe is 
understood as a decrumpling process, much like the one we know from polymeric
surfaces. The dimension  
of space is then a dynamical real variable. The generalized Friedmann
equation, derived from a Lagrangian, and the generalized equation of 
continuity for the matter content of the universe, give the  
dynamics of our model universe. This leads to an oscillatory non-singular 
model with two turning points for the dimension of space.\\ 

\normalsize

\noindent The picture we are proposing for the space-time is a generalization 
of polymeric or tethered surfaces, which are in turn simple 
generalizations of linear polymers to two-dimensionally connected 
networks [1,2]. Visualizing the universe as a piece of paper, then 
the crumpled paper will stand for the state of the early 
universe [3]. As we are not going to develop yet a statistical mechanical 
model for our decrumpling universe, the dynamical model we are   
proposing in this paper could as well be interpreted as a 
generalization of fluid membranes [4]. In this case we can visualize 
the universe as a clay, which can be formed to a three dimensional ball, 
to a two dimensional disc, or even to a one dimensional string. 
In each case the effective dimension of the universe is a continuous 
number between the dimension of the embedding space and some $D_0$ 
which could be 3. To study the crumpling phenomenon in the statistical physics 
one needs to define an embedding space, which does not exist in our 
case. Therefore, we assume an embedding space of arbitrary high 
dimension $\cal D$, which is allowed to be infinite. This is 
necessary, because the crumpling is highly dependent on the 
dimension of embedding space.\\  
Now, imagine the model universe to be  
the time evolution of a D-space embedded in a space with arbitrary large,  
 maybe infinite, dimension $\cal D$. To model the crumpling, we assume the
D-space to be consisted of cells with characteristic size of about the Planck 
length, denoted by $\delta$. The cells, playing the role of the monomers in 
polymerized surfaces, are allowed to have as many dimensions as 
the embedding space. Therefore, the cosmic space can have a dimension 
as large as the embedding space, like the polymers in crumpled phase. 
The radius of gyration of the crumpled cosmic space will play the 
role of the scale factor in a FRW cosmology in 
$D+1$ dimensional space-time, where $D$ is the fractal dimension of 
the crumpled space in the embedding space and could be as high as 
$\cal D$. The expansion of space is understood now as decrumpling of 
cosmic space. In the course of decrumpling the fractal dimension of space  
changes. \\
To formulate the problem we write down the Hilbert-Einstein action  
for a FRW metric in $D$ space dimension. This is along the same line as the 
formulation of homogenous cosmologies using the minisuperspace.  
 Now in our toy model not only the scale factor $a$, but also  
the dimension $D$ of space, are dynamical variables.  
The above mentioned cell structure of the universe brings in 
the next simplification which is a relation between $a$ and $D$.  
It turns out that these 
generalized field equation admits the FRW model as a limit.
For the sake of simplicity we confine ourselves to the flat, $k=0$, case.\\
Let us begin with a $D+1$ dimensional space-time $M\times R$, 
where $M$ is assumed to be homogeneous and isotropic. The space-time 
metric is written as

\be
ds^2=-dt^2+a^2(t)\delta_{ij}dx^idx^j  \qquad i,j=1,...,D  \ee   
The gravitational part of the Lagrangian, assuming $D$ to be a constant, becomes
\be
L_G=-{1\over{2\kappa}}D(D-1)\Big({{\dot a}\over a}\Big)^2a^D,
\ee
where $a^D$ is the volume of $M$, and we have used the homogeneity of the
metric to integrate the Lagrangian density.\\ 
To couple this Lagrangian to the source we use the well-known procedure
in general relativity: write first the matter Lagrangian as 
\be
L_M={1\over 2}\theta^{\mu\nu}g_{\mu\nu},                          \ee
where   
\be
\theta^{00}=\tilde\rho :=\rho a^D,                              \ee  
and
\be
\theta^{ij}=\tilde p\delta^{ij}:=p a^{D-2}\delta^{ij}.
\ee  
Now for the complete Lagrangian we obtain   
\be
L=-{1\over{2\kappa}}D(D-1)\Big({{\dot a}\over a}\Big)^2 a^D+
\Big( -{{\tilde\rho}\over 2}+{{\tilde p D a^2}\over 2}\Big) .
\ee  
We have to vary this Lagrangian with respect to $a$, considering $\tilde \rho$
and $\tilde p$ as constants. Therefore just one equation of motion is obtained.
The continuity equation for the matter content of the cosmological model  
is the other one which we are going to use. It can be shown
that these two equations give us the familiar 2 Friedmann equations. \\ 
To implement the idea of dimension as a dynamical variable, we assume a 
cellular structure for space: the universe consists of $N$ $D_0$-dimensional
cells. To make them eligible to construct higher dimensional configurations, we
assume our cells to have an arbitrary number of extra dimensions, each 
having a characteristic length scale $\delta$. Then the following relation holds
between the $D$- and $D_0$-dimensional volume of the cells:
\be
\mbox{vol}_D(\mbox{cell}) = \mbox{vol}_{D_0}(\mbox{cell})\; \delta^{D-D_0}  
\ee
Now, taking $a$ as the radius of gyration of our decrumpling universe[5], we 
may write
\be
a^D = N\; \mbox{vol}_D (\mbox{cell})\;\delta^{D-D_0}= 
N\; \mbox{vol}_(D_0)(\mbox{cell})\;\delta ^{D-D_0}=
a_0^{D_0} \delta^{D-D_0},
\ee
or
\be
(\frac{a}{\delta})^D= e^C,                            \ee
where $C$ is a constant. Here $D_0$ is a constant dimension which could be
assumed to be 3, and $a_0$ is the corresponding length scale of the 
universe, i.e. the present radius of the universe.\\
Now, we go on to start with the Lagrangian (6), letting the dimension $D$
to be any real number. It is then seen that the Lagrangian (6)  suffers from
the fact that its dimension is not a constant. 
To obtain a Lagrangian with a constant dimension we 
multiply (6) by   
$a_0^{D_0-D}$.  
Now, for our general case of variability of the 
space dimension, the constant part of this factor, $a_0^{D_0}$, can 
be omitted. Therefore, we finally arrive at the Lagrangian[6] 
\be
L=-{{D(D-1)}\over{2\kappa}}\Big(
{{\dot a}\over a}\Big)^2\Big({a\over{a_0}}\Big)^D+\Big( 
-{{\hat\rho}\over 2}+{{\hat p D a^2}\over2}\Big),
\ee   
where
\be
\hat\rho:=\rho\Big({a\over{a_0}}\Big)^D\quad \mbox{, and} \quad   
\hat p:=p a^{-2}\Big({a\over{a_0}}\Big)^D.                             \ee 
Variation of this Lagrangian with respect to $a$ and $D$ leads to a field
equation for $a$ and one for $D$. But there is also the constraint equation
(17). Taking this into account we arrive finally at the equations
\be
(D-1)\left(\frac{\ddot a}{a} + \left[ \frac{D^2}{2D_0}-1-\frac{D(2D-1)}{
2C(D-1)} \right](\frac{\dot a}{a})^2 \right) + \kappa p (1- \frac{D}{2C})  
 = 0   
\ee
and 
\be
a\frac{dD}{da} = - \frac{D^2}{C}.                               \ee 
The field equation (12) is not sufficient to obtain $a$. A 
continuity equation, and an equation of state, are also needed. 
A dimensional reasoning leads to the following generalization of 
the continuity equation [6]:
\be
{d\over{dt}}(\rho a^D a_0^{D_0-D})+p{d\over{dt}}(a^D a_0^{D_0-D})
=0,                                                                  \ee
or
\be
{d\over{dt}}\bigg[\rho\Big({a\over{a_0}}\Big)^D\bigg] 
+p{d\over{dt}}\Big({a\over{a_0}}\Big)^D=0.
\ee     

 The dynamics of our model universe is defined through (12), (13), 
(15), and an equation of state. This is a difficult system to be solved 
analytically.  However, a first integral of motion can be derived which helps
us to understand our model universe qualitatively. It leads to a potential
which can be written for a radiation-like equation of state in the large
$D$ limit in the form   
\be
 U(D)\sim {{D_0}\over{2C}}e^{-C}\Big({D\over D_0}
\Big)^{C\over D_0}.
\ee
Similarly, for $D$ near zero, assuming the pressure to remain 
finite (nonzero), it is obtain to be   
\be
 U(D)\sim -C\ln D,
\ee  
that is, $U$ grows unboundedly to infinity at $D\to\infty$, as 
well as $D\to 0$. It can be shown that the point $D=2C$ is the point where
the potential attains its minimum[6]. This means that there are two 
turning points, one above $D=2C$, the other below it.
The above discussion is valid, provided $T\ge 0$. However, the kinetic term 
(32) changes sign at $D=1$. Therefore, to have two  
turning points, the constant $E:=U+T$ must 
be sufficiently low to make the lower turning point greater than 1. 

A study of the behavior of our model near the lower turning point 
shows that to have any effective change in the space 
dimension one has to go back in time as much as about some multiple of the 
currently assumed age of the universe, and that there is no sensible deviation
from FRW models up to the Planck time.    

\end{document}